# Searching the Internet for evidence of time travelers

## Robert J Nemiroff [1, 2] and Teresa Wilson[1]

[1] Department of Physics, Michigan Technological University, Houghton, MI 49931
Email: nemiroff@mtu.edu

**Abstract.** Time travel has captured the public imagination for much of the past century, but little has been done to actually search for time travelers. Here, three implementations of Internet searches for time travelers are described, all seeking a prescient mention of information not previously available. The first search covered prescient content placed on the Internet, highlighted by a comprehensive search for specific terms in tweets on Twitter. The second search examined prescient inquiries submitted to a search engine, highlighted by a comprehensive search for specific search terms submitted to a popular astronomy web site. The third search involved a request for a direct Internet communication, either by email or tweet, pre-dating to the time of the inquiry. Given practical verifiability concerns, only time travelers from the future were investigated. No time travelers were discovered. Although these negative results do not disprove time travel, given the great reach of the Internet, this search is perhaps the most comprehensive to date.

**Contents**

**1. Introduction**

The origin of the idea of time travel is unknown [1]. Mentions in the distant past include the Indian Mahabharata [2], which may date as far back as the 9th century BC, the Hebrew Talmud [3], written about 300 AD, and the Japanese Nihongi [4], which dates back to about 700 AD. A classic contemporary story of time travel is H. G. Wells' 1895 work "The Time Machine" [5]. All of these, however, predominantly describe time travel to the future. One of the oldest stories known of time travel to the past dates only back to 1733 with Samuel Madden's "Memoirs of the Twentieth Century" [6]. Modern fictional stories involving time travel both to the future and the past are, however, ubiquitous. Two prominent examples include the Doctor Who television series [7] which originated in 1963, and the Back to the Future film series [8] which began in 1985.

---
[2] Author to whom any correspondence should be addressed.

Time travel to the future stands on firm scientific footing. Special Relativity [9] has clear sub-luminal solutions that correspond with time travel to the future. A famous theoretical example is the twin paradox [10]. Such future time travel has been experimentally verified, for example, using a pair of clocks one of which was taken on an airplane. The flying clock recorded a relative time delay of order $10^{-7}$ seconds, in comparison to the more stationary clock [11].

Time travel to the past is controversial, at best, and impossible according to conventional views of the laws of physics. Formally, Special Relativity allows time travel to the past only for objects moving faster than the speed of light [12]. In both Special and General Relativity, the possibility that objects could travel on closed timelike loops would indicate the possibility of time travel to the past. Solutions involving closed timelike loops have been found in General Relativity with popular cases including Gödel's universe [13], traveling between black holes and wormholes [14], and circling cosmic strings [15]. Many physicists consider such solutions unphysical, as articulated, for example, by the Chronology Protection Conjecture [16].

Although less well known than popular fiction, experiments designed to discover human time travelers have been conducted. In May of 2005, then graduate student A. Dorai at MIT publicised and held a convention for time travelers [17]. No one claiming to come from the future showed up [17]. S. Hawking did a similar experiment in July of 2012, holding a personal party for time travelers, but sending out the invitations only after the party [18]. No one claiming to be a time traveler showed up [18].

In this work, we report on a series of searches for digital signatures that time travelers potentially left on the Internet. Specifically, we search for content that should not have been known at the time it was posted. Such information is here referred to as "prescient". To the best of our knowledge, no similar search has ever been published previously. Section 2 of this work outlines possible types of time travelers. Section 3 describes one search for prescient content placed on the Internet, highlighted by a search of tweets on Twitter. Section 4 describes a search for prescient terms submitted in Internet search engines, highlighted by a search for two specific terms in the online search engine for the Astronomy Picture of the Day web site. Section 5 describes an experiment involving the request for a prescient timed communication to be sent either as a tweet or an email. Section 6 summarizes the results and draws some conclusions concerning the nature of the results.

## 2. Types of Time Travelers

Time travelers can be classified as either from the past or from the future. Further subdivisions might distinguish between time travelers who want to advertise their presence, hide their presence, and those who are indifferent. We do not consider such distinctions as determinative for search methods employed here, however, as even time travelers who want to advertise their presence may do so ineffectively, those who want to hide their presence might make a revealing mistake, and those indifferent might or might not leave traceable Internet content.

We did not search the Internet for evidence of time travelers from the past. First, we were unable to conceive of a simple method that would clearly indicate that informational traces they might have left were evidence of time travel from the past and not just simple knowledge of the past. Next, to the best of our knowledge, human technology to create a time machine does not exist in the past, so that time travelers from the past must originate in the future, assuming such technology is ever developed.

## 3. Searching for Prescient Content on the Internet

Were a time traveler from the future to access the Internet of the past few years, they might have left once-prescient content that persists today. Alternatively, such information might have been placed on Internet by a third party discussing something unusual they have heard. Such content might have been catalogued by search engines such as Google (google.com) or Bing (bing.com), or remain in posts left on Facebook (facebook.com), Google Plus (plus.google.com), or Twitter (twitter.com).

To make our search for such content manageable, a small number of search terms were sought that could conceivably give a relatively clear signal of prescient information in an automated query of a large database. The time period of our search was from 2006 January to 2013 September. Search terms were chosen to optimize three major attributes. First, search terms were sought that described an item or event that acquired a name during the time period of our search. Such terms would allow two different epochs to be compared -- potentially prescient content before this information became available, and normal content afterwards. For example, mentions of the "War of the Roses" occurred numerous times, but none that could easily be tagged in an automated search as even potentially prescient, since the actual Wars of the Roses ended in 1485, and the name "Wars of the Roses" only came into common usage in the 1800s, well before our automated searches began [19]. Therefore all discussions in our search period could be attributed to normal discussions of historic events.

Second, search terms should involve as unique a label as possible. Were a large automated search focused on a common term, it would be difficult to automatically differentiate a prescient mention of this term from likely much more numerous common mentions. For example "Comet McNaught" was not considered to be a good search term because its discoverer, Robert H. McNaught, has discovered over 50 comets dating as far back as 1987. Therefore, an automated search returning mentions of "Comet McNaught" in 2006, for example, could arguably not be returning prescient mentions of the impressively bright Comet McNaught in 2009, but rather mentions of one of the earlier-discovered comets that bears the same popular name.

Third and last, search terms should remain well known and historically important into the future. Perennially famous objects or events might be more likely to be on the agendas of future time travelers, and so may permeate their Internet interactions. Not being time travelers ourselves, we cannot know for sure what present-day labels will remain popular into the future, but focusing on modern renditions of terms used by historically long-standing and internationally known institutions seemed pragmatic. Based on these criteria, two main labels were chosen: Comet ISON and Pope Francis.

Comet ISON (C/2012 S1) is a comet that was discovered by the International Scientific Optical Network (ISON) on 2012 September 21. Therefore, the term "Comet ISON" came into the public lexicon during our search window. Although the ISON network had discovered one previous comet, that comet did not go by the popular name of Comet ISON. Therefore, the term "Comet ISON" is a unique label that should not be easily confused with something else. Comet ISON is internationally known and has been a topic of popular discussion on the Internet since its discovery. Histories of bright comets like Comet ISON are generally well kept by societies and journals around the world, indicating that Comet ISON might remain memorable well into the future. Conversely, there is little reason for anyone without prescient information to be referring to something as "Comet ISON" before 2012 September. Therefore, discussions or even mentions of "Comet ISON" before 2012 September were searched for as potentially prescient evidence of time travelers from the future.

On 2013 March 16, the newly elected pope of the Catholic Church, Jorge Mario Bergoglio, chose the official name of Francis. Bergoglio is the first pope ever to choose the name Francis. Therefore, the term "Pope Francis" is relatively unique and came into the public lexicon during our search period. Since Christianity is currently the most popular religion on Earth, Roman Catholics comprise the largest sect of Christianity [20], and papal histories are well recorded, it seems reasonable to assume that "Pope Francis" would remain memorable well into the future. Before 2013 March, however, there is little reason for anyone without prescient information to mention a "Pope Francis".  Discussions or even mentions on the Internet of Pope Francis before 2013 March were therefore searched for as potentially prescient evidence of time travelers from the future.

A relatively efficient method for locating specific content on the voluminous Internet is through hashtags. Hashtagging -- labeling Internet content with terms beginning with a "#" --  originated on the Internet in the Internet Relay Chat (IRC) channels, which itself began in 1988 [21]. Hashtagging became prevalent on Twitter in 2007 as a means of signaling that information about given topics is available [22]. In the past few years, hashtagging has become increasingly prevalent beyond Twitter and across the Internet, being used frequently on Facebook, Google Plus, blogs, and general Internet content. In essence, hashtags make information easier to find in online searches. We therefore typically included the hashtagged terms "#cometison" and "#popefrancis" in our searches. For example, when someone places content involving Comet ISON on the Internet, they might include the hashtag "#cometison". Since that hashtagged term would not occur in colloquial text, it acts as a label that can be found in a search by people seeking information about Comet ISON. Conversely, labeling and searching for the hashtagged term "#comet" or "#ISON" might not return information about only Comet ISON, but typically voluminous content involving other comets or other instances of the term "ison", a word that means "great" in Finnish.

A direct way to search for information posted on the web is to use a commonly popular search engine. At the time of this writing, the most popular search engine is Google. Unfortunately, searching for prescient web content using the publicly available Google search has proven unreliable. Although providing the ability to sort identified content by date, several exploratory tests on Google found an initially surprising number of web pages that contained seemingly prescient information. Upon further inspection, however, all potentially-prescient content on those web pages was clearly non-prescient. One prominent reason for this was the appearance of recent advertisements on older news stories. Bing

(bing.com), another currently popular search engine, did not appear to have a sufficient ability to filter results by posting date to be useful.

Another place potentially prescient content might be found is on posts to the social network Facebook. At the time of this writing, Facebook is the most popular social network [23]. Facebook pages incorporate their own search engine. Unfortunately, searches on Facebook turned up results that were clearly not comprehensive. Searches for prescient posts on Facebook frequently culminated in an unexpectedly short listing, with text at the bottom stating "There are no more recent posts to show right now." The word "recent" was taken to indicate that the Facebook search did not report older, potentially prescient posts. Additionally, Facebook also allows backdated posts, as far back as the day the Facebook account was created. Such backdating undermines the ability of a search to determine whether information posted to Facebook is prescient. For example, someone who opened a Facebook account in 2006 could have posted a comment about "Pope Francis" in 2013 April and then backdated that post to 2008 April. In a search, we might discover this post and wrongly consider it prescient knowledge in 2008 that a "Pope Francis" would exist in the then-future 2013. However, even given Facebook's limited utility, searches found no prescient mentions of "Comet ISON", "Pope Francis", "#cometison", or "#popefrancis".

Similarly, a preliminary search for prescient content on the social network Google Plus (plus.google.com) was terminated after it became clear that Google Plus did not always order search results temporally. It was therefore too difficult, in practice, to find older and potentially prescient informational mentions.

Our most comprehensive search for potentially prescient Internet content was achieved using the microblogging Internet platform Twitter [24]. Twitter enables members to "tweet" posts up to 140 characters. Twitter was created in 2006 and has been growing in popularity ever since. As of 2013 May, over 500 million active registered Twitter accounts produce over 1 billion tweets per month [24]. Even so, given the limited number of characters per tweet, a search using Twitter's own search engine appeared quick, comprehensive, had the ability to find tweets back to the very beginning of Twitter, and order the tweets temporally. Exploratory searches on Twitter indicated the ability to find all tweets that mentioned any of our preferred search terms. Also, Twitter does not allow backdated tweets. We therefore considered our search on Twitter to be our most comprehensive search for once-prescient content placed on the Internet.

No clearly prescient content involving "Comet ISON", "#cometison", "Pope Francis", or "#popefrancis" was found from any Twitter tweet -- ever. One candidate was found -- an interesting speculative discussion using the term "Pope Francis" in a blog post advertised by a tweet, but upon close inspection and consideration, that blog post was deemed overtly speculative and not prescient. Searches were made both on the embedded Twitter search box and on the Twitter search service Topsy (topsy.com). Each of these search terms occurred numerous times -- hundreds for Comet ISON and thousands for Pope Francis -- but, with the one noted exception, only after 2012 September for Comet ISON and only after 2013 March for Pope Francis.

Note that information once posted to the Internet may be subject to subsequent deletion by the poster, including content on the web, Facebook, and Twitter. Therefore, in general, information that was searched for potentially prescient content is that which remained on the Internet in August 2013.

**4. Searching for Prescient Search Queries**

Another type of informational trace a time traveler from the future might have left on the modern-day Internet is a query to a current search engine for not-yet-available information. For example, a time traveler might have been trying to collect historical information that did not survive into the future, or might have searched for a prescient term because they erroneously thought that a given event had already occurred, or searched to see whether a given event was yet to occur. In an effort to uncover such prescient queries, we searched online databases for potentially prescient search terms themselves.

Fortunately, searches using Google, the most popular search engine of the present time, are themselves searchable with Google Trends (google.com/trends) [25]. Google Trends was therefore used for a search for potentially prescient queries involving Comet Ison and Pope Francis. Although numerous searches were uncovered, none occurred sufficiently early to be considered prescient. Unfortunately however, we do not consider these negative results reliable. One reason is that the publicly available Google Trends only reported search results that had a significant "search volume" [25]. Therefore, for example, Google Trends did not report any instances of the term "#cometison" being searched for -- ever. And since we searched for this term ourselves, we know that at least some instances did occur. Next, Google Trends only reported on the prevalence of searches as normalized to the largest search volume in the desired time window, and not in absolute terms. Therefore, for example, search reports on "Comet ISON" reported a zero score for all days from January 2004 through September 2012, the month that Comet ISON was discovered, but numerous search queries thereafter. This zero score, however, was normalized to the peak score set to 100 for 2013 March. The raw numbers of searches for March 2013 were not revealed by Google Trends. Therefore, to our understanding, the zero score really meant "less than 0.5 percent of the March 2013 value", which could well be greater than zero. Quite possibly, a single prescient search for Comet ISON would not have been recorded. Similarly Google Trends searches for the term "Pope Francis" gave a zero score for all times before 2013 March, and numerous searches thereafter, but was also not considered reliable. Neither Bing nor Yahoo! currently provide a public feature that allows a search over past queries.

A search that was sensitive to a single potentially prescient query was in the log files for the primary internal search engine affiliated with the Astronomy Picture of the Day (APOD) web site (apod.nasa.gov), as housed at the USA's National Air and Space Administration (NASA) [26]. One of the authors (RJN) is an APOD site editor and is able to access daily log files which include queries sent to APOD's internal search engine. The inspected files did not contain personal information, and no attempts were made to connect any search information to any specific people.

The NASA APOD web page is served with Apache software running under the Linux operating system. The search page used the URL apod.nasa.gov/cgi-bin/apod/apod_search. This page returned a search form that utilized the open source and free search software titled Simple Web Indexing System for

Humans - Enhanced (Swish-e) [27]. For many years, including the time between 2006 January and 2009 March, the NASA APOD search page HTML code used the "isindex" element to record and send back search terms to the web server at NASA's Goddard Space Flight Center. This HTML code returned the search page URL appended with a suffix of a question mark followed by the search term to an indexed database previously catalogued by Swish-e. The returned query line was then indexed in a log file. For example, a search for "M31", without quotes, would result in the log file writing a line with the text "?M31" in it, without quotes.

The NASA APOD log files were queried for potentially prescient search terms from the beginning of 2006 January through the end of 2009 March. Specifically, given that the Linux environment where the search was done was case sensitive, a search for either "?ISON", "?Ison", or "?ison" was conducted in the daily log files, quotes excluded. To make sure our search procedure was working, a concurrent search for the non-prescient term "?M31" was also conducted. M31 is the Messier catalog number for the Andromeda galaxy, a commonly used designation for a well-known astronomical object. The search for the "?M31" term uncovered many instances, typically several for each day searched, whereas only a handful of instances of any of the terms "?ISON", "?Ison", or "?ison" were ever found in the log files during this time period. Upon further investigation, each of these instances was related to extraneous information or misspellings, leaving no queries as possibly prescient.

Starting in 2009 April, the way the NASA APOD search engine worked was changed so that queries submitted through the main APOD search page no longer typically recorded search results with a "?" prefix. Therefore, at that time, our search for prescient terms after this date could be incomplete, and potentially prescient searches missed.

**5. Requests for Time Travelers to Issue a Prescient Internet Communication**

In addition to searching for prescient information that time travelers might have left archived *passively* on the Internet, another approach used was to make a request for time travelers to *actively* respond to a request for a prescient communication. The main idea was not to converse with individual time travelers, but rather to encourage a single interchange indicating that time travel has become possible in the future. The versatility of this approach allowed for the inclusion of additional information about whether time travel can alter these authors' past.

An experiment designed to encourage such active communication was structured as follows. A post was created in 2013 September on a publicly available online bulletin board requesting that one of two hashtags be tweeted or emailed before a certain date. Specifically, time travelers were requested to respond with a communication including either the hashtagged term "#ICanChangeThePast2" or "#ICannotChangeThePast2" on or before 2013 August.

A message incorporating the hashtagged term "#ICannotChangeThePast2" would indicate that time travel to the past is possible but that the time traveler believes that they do not have the ability to alter the authors' past. A universe where the past cannot be changed is termed as having a "fixed history", where history can be regarded as a single timeline. Such universes may uphold the Novikov Self-Consistency

Conjecture [28]. For example, in a fixed history universe, nothing the time traveler could do would change the existence of the Wars of the Roses, so content involving that event could always be found by the authors on the Internet.

Conversely, a message incorporating the hashtagged term "#ICanChangeThePast2" would indicate that time travel to the past is possible and that the time traveler can demonstrate the ability to alter the authors' past. Theoretical universes where the past can be changed are termed as having a "plastic history", where history cannot be regarded as a single timeline [29]. For example, in a plastic history universe, a time traveler might have the ability to go back and change history so that the Wars of the Roses never occurred in the past of the authors, with the result that the authors would not be able to find content about this event on the Internet. Time travel in a plastic history universe could introduce seeming logical conundrums such as the "grandfather paradox" [30], where the time traveler's grandfather, for example, could be killed young enough to sabotage the very existence of the time traveler. It is not the purpose of this section to detail, debate, or resolve logical conundrums convolved with plastic histories -- just to carry out a simple test for them.

In 2013 August we searched all of Twitter for the hashtag "#ICanChangeThePast2" and found no occurrences. Starting in 2013 September, time travelers were publically requested to go into their past and tweet "#ICanChangeThePast2" before 2013 August to demonstrate this ability. After this request, later in 2013 September, we again searched Twitter for the same hashtag, but only for a tweet with a posting date before 2013 August, as requested. For such a tweet to have been found only in the second search and not the first, a time traveler would have had to have known about the advertisement posted in 2013 September, go back in time to before 2013 August, send the prescient tweet, and have this tweet appear only in our later Twitter search in 2013 September. Assuming no spoofing, the only way this can happen is if time travelers can change our past with their actions.

Similarly, in 2013 August, we searched the cataloged email sent to "home.nemiroff@yahoo.com" for a communication with the term "#ICanChangeThePast2" in the subject line and found no occurrences. Starting in 2013 September, time travelers were publically requested to go into their past and send an email with "#ICanChangeThePast2" in the subject line to "home.nemiroff@yahoo.com" before anytime between 2008 November, the month after this email address was created, and 2013 August.

Alternatively, time travelers who believe that they cannot change the past were requested to tweet "#ICannotChangeThePast2" on or before 2013 August, or include that hashtag at the end of an email sent to "home.nemiroff@yahoo.com" between 2008 November and 2013 August. For this part of the experiment, we were careful NOT to search Twitter or email until after the hashtag was publicly advertised in 2013 September. Therefore, by not pursuing a pre-advertisement search, the past might not have had to be changed in order for this tweet to have been discovered in 2013 September.

Advantages of using Twitter for this active search include that tweets are discoverable by anyone with sufficient access to the Internet and so are publically falsifiable. Advantages to using email include the relative ease of acquiring even a temporary email account, and the relative amount of privacy -- perhaps preferred by the time traveler -- inherent in a private email message.

Unfortunately, as of this writing, no prescient tweets or emails were received. Given the additional exposure that the public listing of this manuscript gains, we will continue to search, on occasion, for active tweets and emails involving potential time travel.

**6. Summary and Conclusions**

The discovery of time travel into the past could be transformative not only to physics but humanity. Its uses might be many and varied -- probably enlightening, but possibly catastrophic. Unlike modern searches for dark matter and dark energy, however, modern searches for time travel are comparatively rare. Given the modern ubiquity of the Internet, though, there now exists new, far reaching, and falsifiable methods to search for time travelers.

Technically, what was searched for here was not physical time travelers themselves, but rather informational traces left by them. Although such information might be left by physical time travelers, conceivably only information itself could be sent back in time, which would be a type of time travel that might not directly involve the backwards transport of a significant amount of energy or momentum. This might be considered, by some, a more palatable mode of backwards time travel than transferring significant amounts of matter or energy back in time, as the later might break, quite coarsely, local conservation of energy and momentum [31]. For example, were the same person at different epochs to stand next to themselves, the energy tied into their own rest mass seems not to have been conserved. Similarly, instantaneous time travel to the same place on Earth might violate conservation of momentum, as the motion of the Earth around the Sun (etc.) might delegate a significant change in momentum for a corporeal object even over a time scale of minutes.

Although the negative results reported here may indicate that time travelers from the future are not among us and cannot communicate with us over the modern day Internet, they are by no means proof. There are many reasons for this. First, it may be physically impossible for time travelers to leave any lasting remnants of their stay in the past, including even non-corporeal informational remnants on the Internet. Next, it may be physically impossible for us to find such information as that would violate some yet-unknown law of physics, possibly similar to the Chronology Protection Conjecture [16]. Furthermore, time travelers may not want to be found, and may be good at covering their tracks. Additionally, time travelers just may not have left the specific event tags that we were searching for. Finally, our searches were not comprehensive, so that even if time travelers left the exact event tags searched for here, we might have missed them due to human error, oversight, incompleteness of Internet catalogs and searches, or inaccurate content time tags.

Nevertheless, given the current prevalence of the Internet, its numerous portals around the globe, and its numerous uses in communication, this search might be considered the most sensitive and comprehensive search yet for time travel from the future.

**Acknowledgements**

We thank Noah Brosch, Abhilash Kantamneni, Jerry Bonnell and members of the online bulletin board The Asterisk for helpful discussions. We also thank Stephen Fantasia for help in understanding the APOD log files and Andrew Robare for suggesting "Pope Francis" as the basis for useful search terms.